\documentclass[twocolumn,showpacs,amsmath,amssymb,10pt]{revtex4}
\usepackage{amsfonts}
\usepackage{bm}

\newcommand{\ot}{{\,\otimes\,}}
\newcommand{{\Cd}}{{\mathbb{C}^d}}

\def\oper{{\mathchoice{\rm 1\mskip-4mu l}{\rm 1\mskip-4mu l}%
{\rm 1\mskip-4.5mu l}{\rm 1\mskip-5mu l}}}
\def\<{\langle}
\def\>{\rangle}
\newtheorem{theorem}{Theorem}

\newtheorem{proposition}{Proposition}
\newtheorem{example}{Example}

\begin{document}
\title{\textbf{Bell diagonal states with maximal abelian symmetry}}
\author{Dariusz Chru\'sci\'nski and Andrzej Kossakowski\\
Institute of Physics, Nicolaus Copernicus University \\
Grudzi\c{a}dzka 5/7, 87--100 Toru\'n, Poland }


\begin{abstract}
We provide a simple class of 2-qudit states for which one is able to
formulate necessary and sufficient conditions for separability. As a
byproduct we  generalize well known construction provided by
Horodecki et al. for $d=3$. It is hoped that these states with known
separability/entanglement properties may be used to test various
notions in  entanglement theory.

\end{abstract}

\maketitle

\section{Introduction}

Physicists love solvable models, that is, models (or problems) for
which one is able to answer all interesting questions. Clearly,
solvable models are not generic. They are rather exceptional.
However, they provide examples where one can study basic properties
of generic models and to test various important notions. Usually,
the `solvability' or `integrability' of the model is connected with
the existence of the special symmetry.  In many cases the presence
of the symmetry  enables one to simplify the analysis of the
corresponding problems and very often it leads to much deeper
understanding and the most elegant mathematical formulation of the
corresponding physical theory.

In quantum information theory \cite{QIT,HHHH} the idea of symmetry
was first applied by Werner \cite{Werner1} to construct an important
family of bipartite $\mathbb{C}^d \ot \Cd$ quantum states which are
invariant under the following local unitary operations: $\rho
\rightarrow U \ot U \rho (U \ot U)^\dagger$,  for any $U\in U(d)$ ,
where $U(d)$ denotes the group of unitary $d \times d$ matrices.
Another family of symmetric states -- so called isotropic states --
\cite{Horodecki} is governed by the following invariance rule $\rho
= U \ot \overline{U} \rho (U \ot \overline{U})^\dagger$, where
$\overline{U}$ is the complex conjugate of $U$ in some fixed basis
in $\Cd$. Both families are `solvable', that is, one can easily
check wether a given  $U \ot U$-- or $U \ot \overline{U}$--invariant
state is separable or entangled.

In this paper we provide a class of states of a quantum composed
system living in $\mathbb{C}^d \ot \mathbb{C}^d$ for which one can
easily check wether a given state is separable or entangled. It is
well known that in general this problem is very hard
\cite{HHHH,Guhne}.

Our construction presented in the next section contains two
ingredients:

1) we restrict to Bell diagonal states
\cite{Bell-2,Bell-3,Bell-4,Werner}, and

2) we assume that $\rho$ is invariant under the action of the local
group $U \ot \overline{U}$, where $U$ belongs to the maximal
commutative subgroup of $U(d)$ \cite{PPT-nasza}.

 As a byproduct we provide a generalization of `solvable' class of
states constructed for $d=3$ by Horodecki et al. \cite{HOR}. It is
hoped that our `solvable' family of states finds applications in
Quantum Information Theory.

\section{Definition of the `solvable' class}

Let $\{|0\>,\ldots,|d-1\>\}$ denotes an orthonormal basis in
$\mathbb{C}^d$ and let $S : \mathbb{C}^d \rightarrow \mathbb{C}^d$
be a shift operator defined by
\begin{equation}\label{}
    S|k\> = |k+1\> \ ,\ \ \ ({\rm mod}\ d) \ .
\end{equation}
Consider now a simplex of Bell diagonal states
\cite{Bell-2,Bell-3,Bell-4} defined by (actually, one may provide
more general definition, cf. \cite{Werner})
\begin{equation}\label{Bell}
    \rho = \sum_{m,n=0}^{d-1} p_{mn} P_{mn}\ ,
\end{equation}
where $p_{mn}\geq 0$, $\ \sum_{m,n}p_{mn}=1$ and
\begin{equation}\label{}
    P_{mn} = (\mathbb{I} \ot U_{mn}) \,P^+_d\, (\mathbb{I} \ot U_{mn}^\dagger)\ ,
\end{equation}
with $U_{mn}$ being the collection of $d^2$ unitary matrices defined
as follows
\begin{equation}\label{U_mn}
    U_{mn} |k\> = \lambda^{mk} S^n |k\> = \lambda^{mk} |k+n\>\ ,
\end{equation}
where
\begin{equation}\label{}
    \lambda= e^{2\pi i/d} \ ,
\end{equation}
and
\begin{equation}\label{}
P_{00} \equiv P^+_d = \frac 1d \sum_{i,j=0}^{d-1} |ii\>\<jj| \ ,
\end{equation}
denotes canonical maximally entangled state in $\Cd \ot \Cd$.
Actually, Bell diagonal states (\ref{Bell}) belong to much more
general class called {\em circulant states} \cite{CIRCULANT} (see
also \cite{Art}). Let us define
\begin{equation}\label{}
    \Pi_0 = \frac 1d\, \sum_{i=0}^{d-1} |ii\>\<ii|\ ,
\end{equation}
and
\begin{eqnarray}\label{}
    \Pi_n &=& (\mathbb{I} \ot S^n) \, \Pi_0 \, (\mathbb{I} \ot S^n)^\dagger
    \nonumber \\ &=&
     \frac 1d\, \sum_{i=0}^{d-1} |i,i+n\>\<i,i+n|\ ,
\end{eqnarray}
for $n=1,\ldots,d-1$. One has
\begin{equation}\label{I}
\Pi_m \Pi_n = \frac 1d\, \delta_{mn} \Pi_n\ ,
\end{equation}
together with
\begin{equation}\label{II}
    \Pi_0 + \Pi_1 + \ldots + \Pi_{d-1} = \frac 1d\, \mathbb{I}_d \ot
    \mathbb{I}_d\ .
\end{equation}

\begin{proposition}[\cite{PPT-nasza}]
Any state $\rho$ defined by
\begin{equation}\label{FAM-UU}
    \rho = \sum_{m=0}^{d-1}  \Big[ \mu_m \Pi_m +  \nu_m P_{m0}\Big]
    \     ,
\end{equation}
 satisfies
\begin{equation}\label{UU}
    U_\mathbf{x} \ot \overline{U}_\mathbf{x} \, \rho = \rho\, U_\mathbf{x} \ot \overline{U}_\mathbf{x}\ ,
\end{equation}
where
\begin{equation}\label{}
U_\mathbf{x} = \exp\left(i \sum_{k=0}^{d-1} x_k |k\>\<k| \right)\ ,
\end{equation}
and $\mathbf{x}=(x_0,\ldots,x_{d-1}) \in \mathbb{R}^d$.
\end{proposition}
Consider now the following family of states
\begin{equation}\label{FAM}
    \rho = \sum_{i=1}^{d-1}\lambda_i \Pi_i + \lambda_d P^+_d\ .
\end{equation}
Note that (\ref{FAM}) defines a subclass of (\ref{FAM-UU}).  One has
$\lambda_1,\ldots,\lambda_d\geq 0$, and $
\lambda_1+\ldots+\lambda_d=1$. Clearly, the family of states
(\ref{FAM}) defines $(d-1)$-dimensional simplex with vertices
$\Pi_1,\ldots,\Pi_{d-1}$ and $P^+_d$. One has the following

Note, that $\Pi_1,\ldots,\Pi_{d-1}$ define separable states (they
are diagonal in the product basis $|ij\> = |i \ot j\>$). The family
(\ref{FAM}) can be fully characterized due to the following

\begin{theorem}
A state $\rho$ is {\rm PPT} if and only if
\begin{equation}\label{PPT}
    \lambda_i \lambda_{d-i}\geq \lambda_d^2\ .
\end{equation}
Moreover, a state $\rho$ is separable if and only if
\begin{equation}\label{SEP}
    \lambda_i \geq \lambda_d\ ,
\end{equation}
for $i=1,\ldots,d-1$.
\end{theorem}
The proof of PPT condition (\ref{PPT}) is easy and it is already
contained in \cite{PPT-nasza}. In the present paper we provide the
proof of separability condition (\ref{SEP}). Suppose that condition
(\ref{SEP}) is satisfied. Any state from the family (\ref{FAM}) can
be represented as the following convex combination
\begin{equation}\label{FAM-1}
    \rho = d\lambda_d\, \widetilde{\rho} + \sum_{i=1}^{d-1}(\lambda_i - \lambda_d) \Pi_i \ ,
\end{equation}
where
\begin{equation}\label{tilde}
\widetilde{\rho} = \frac 1d \left( \sum_{i=1}^{d-1} \Pi_i +
P^+_d\right)\ .
\end{equation}
Now, it is well known that $\widetilde{\rho}$ is separable. Hence
$\rho$ is separable being the convex combination of separable states
$\widetilde{\rho}$ and $\Pi_0,\Pi_1,\ldots,\Pi_{d-1}$. To prove that
separability implies (\ref{SEP}) one needs to devise an appropriate
entanglement witness.

\section{Entanglement witnesses}

To define a border between separable and entangled states one needs
an appropriate family of entanglement witnesses. Let us recall
\cite{HHH,Terhal1,Terhal2} (see laso \cite{O,Lew2,Lew3}) that a
Hermitian operator $W$ defined on a tensor product
$\mathcal{H}=\mathcal{H}_1 \ot \mathcal{H}_2$ is called  an EW iff
1) $\mbox{Tr}(W\sigma_{\rm sep})\geq 0$ for all separable states
$\sigma_{\rm sep}$, and 2) there exists an entangled state $\rho$
such that $\mbox{Tr}(W\rho)<0$ (one says that $\rho$ is detected by
$W$). It turns out that a state is entangled if and only if it is
detected by some EW \cite{HHH}. The simplest way to construct EW is
to define $W = P + Q^\Gamma$, where $P$ and $Q$ are positive
operators. It is easy to see that $\mbox{Tr}(W\sigma_{\rm sep})\geq
0$ for all separable states $\sigma_{\rm sep}$, and hence if $W$ is
non-positive, then it is EW. Such EWs are said to be decomposable
\cite{O}. Note, however, that decomposable EW cannot detect PPT
entangled state  and, therefore, such EWs are useless in the search
for bound entangled state. An EW which is not decomposable is called
indecomposable (or non-decomposable). A PPT state $\rho$ is
entangled iff there exists an indecomposable EW such that ${\rm
Tr}(\rho W)<0$.

Let us consider the following family of Hermitian operators
\begin{equation}\label{W-dk}
    W_{d,k} = (d-k) \Pi_0 + \sum_{i=1}^{k}
    \Pi_i - P^+_d \ ,
\end{equation}
for $k=1,2,\ldots,d-1$. It is well known \cite{Osaka1,Osaka2} (see
also \cite{Chr,How}) that for $k=1,\ldots,d-2\,$, $W_{d,k}$ defines
an indecomposable EW, and $W_{d,d-1}$ is a decomposable EW.
Actually, $W_{d,d-1}^\Gamma \geq 0$ and it corresponds to the
reduction map, that is
\begin{equation}\label{}
    W_{d,d-1} = (\oper \ot R)P^+_d\ ,
\end{equation}
where  $R(X) = \mathbb{I}_d {\rm Tr}\, X - X$. Now, it is evident
from (\ref{I}) and (\ref{II}) that that the role of normalized
projectors $\Pi_1,\ldots,\Pi_{d-1}$ is perfectly symmetric. Hence,
for any permutation
\begin{equation}\label{}
    \pi : \{0,1,\ldots,d-1\} \longrightarrow
    \{\pi(1),\ldots,\pi(d-1)\}\ ,
\end{equation}
the new operator
\begin{equation}\label{W-dk}
    W^\pi_{d,k} = (d-k) \Pi_0 + \sum_{i=1}^{k}
    \Pi_{\pi(i)} - P^+_d \ ,
\end{equation}
is again the legitimate EW. Note, however, that the property of
(in)decomposability  is not preserved for an arbitrary permutation,
that is, $W^\pi_{d,k}$ might be decomposable/indecomposable even if
$W_{d,k}$ is indecomposable/decomposable.

Now,  if $\rho$ defined in (\ref{FAM}) is separable then ${\rm
Tr}(\rho W_{d,1}^\pi) \geq 0$. One has
\begin{equation}\label{}
    {\rm Tr}( \rho\, W_{d,1} ) = \frac 1d (\lambda_1 - \lambda_d) \ ,
\end{equation}
and hence, separability of $\rho$ implies $\lambda_1 \geq
\lambda_d$. Taking an arbitrary permutation $\pi$ such that $\pi(1)
= k$ one finds
\begin{equation}\label{}
    {\rm Tr}( \rho\, W^\pi_{d,1}) = \frac 1d (\lambda_k - \lambda_d) \ ,
\end{equation}
which finally proves (\ref{SEP}).

\begin{example}{\em Consider  a state $\rho$ defined in (\ref{FAM})
with
\begin{equation}\label{epsilon}
    \lambda_1 = \frac{\varepsilon}{N_\varepsilon}\ , \ \ \lambda_{d-1} =
    \frac{1}{\varepsilon N_\varepsilon} \ ,
\end{equation}
and
\begin{equation}\label{}
\lambda_2 = \ldots =\lambda_{d-2} = \lambda_d =
\frac{1}{N_\varepsilon}\ ,
\end{equation}
 where $\varepsilon>0\,$, and the normalization factor
\begin{equation}\label{}
    N_\varepsilon = d(d-2 + \varepsilon + \varepsilon^{-1})\ .
\end{equation}
One has $\lambda_1 \lambda_{d-1} = \lambda_d^2$ which shows that
$\rho$ is PPT for all $\varepsilon>0$. However, the separability
condition (\ref{SEP}) is not satisfied unless $\varepsilon=1$.
Hence, for $\varepsilon \neq 1$ a state (\ref{epsilon}) is PPT but
entangled. Note that for $\varepsilon=1$  one has $\rho =
\widetilde{\rho}\,$, where $\widetilde{\rho}$ is defined in
(\ref{tilde}). }
\end{example}

\begin{example}{\em
Consider now the special case of (\ref{FAM}) defined by
$$ \lambda_2 = \ldots =\lambda_{d-2} = \lambda_d\ , $$
where
\begin{eqnarray} \label{lll}
  \lambda_1 &=& \frac{\alpha}{N} \ , \nonumber \\
  \lambda_{d-1} &=&  \frac{(d-1)^2+1 - \alpha}{N} \ ,  \\
  \lambda_d &=&  \frac{d-1}{N} \ , \nonumber
\end{eqnarray}
with
\begin{equation}\label{}
    N = (d-1)(2d-3) +1\ .
\end{equation}
The parameter $\alpha \in [0,(d-1)^2+1]$. It is clear that for $d=3$
one recovers Horodecki construction \cite{HOR}: $\Pi_1=\sigma_+$,
$\Pi_2=\sigma_-$, and
\begin{equation}\label{d=3}
    \rho_\alpha = \frac 27\, P^+_3 + \frac{\alpha}{7}\, \sigma_+ +
    \frac{5-\alpha}{7}\, \sigma_-\ ,
\end{equation}
with $\alpha\in [0,5]$.  It is well known that a state (\ref{d=3})
is PPT for $\alpha \in [1,4]$. Moreover, it is separable for $\alpha
\in [2,3]$. Hence, for $\alpha \in [1,2) \cup (3,4]$ it is PPT
entangled. Now, we perform a similar analysis for a generalized
state. A state (\ref{FAM}) defined by (\ref{lll})  is PPT if and
only if
   $ \lambda_1 \lambda_{d-1} \geq \lambda_d^2\,$
which implies
\begin{equation}\label{beta}
    1 \leq \alpha \leq (d-1)^2\ .
\end{equation}
Hence $\rho$ is separable if and only if $\lambda_1,\lambda_{d-1}
\geq \lambda_d\,$ which is equivalent to
\begin{equation}\label{}
  d-1 \leq   \alpha \leq (d-1)(d-2)+1 \ .
\end{equation}
Hence, for
\begin{equation}\label{}
    \alpha \in [1,d-1) \, \cup\, ((d-1)(d-2)+1,(d-1)^2]\ ,
\end{equation}
a state is PPT but entangled. }
\end{example}

\section{Generalized isotropic states}

Consider now a simple generalization of (\ref{FAM}) provided by
\begin{equation}\label{FAM-G}
    \rho = \sum_{i=0}^{d-1}\lambda_i \Pi_i + \lambda_d P^+_d\
\end{equation}
that is, one adds an additional term `$\lambda_0 \Pi_0$'. We stress
that (\ref{FAM-G}) still satisfies (\ref{UU}). However, in general
Theorem 1 is no longer true. It is clear that  PPT condition does
not change: a state (\ref{FAM-G}) is PPT iff the condition
(\ref{PPT}) is satisfied. Note that (\ref{SEP}) implies
separability. Indeed, one has
\begin{equation}\label{FAM-1}
    \rho = d\lambda_d\, \widetilde{\rho} + \lambda_0 \Pi_0 +
    \sum_{i=1}^{d-1}(\lambda_i - \lambda_d) \Pi_i \ ,
\end{equation}
where $\widetilde{\rho}$ is defined in (\ref{tilde}). Hence, it is a
convex combination of separable states. Note however that now
condition (\ref{SEP}) is only sufficient but not necessary for
separability of (\ref{FAM-G}).  The necessary conditions ${\rm
Tr}(\rho\, W^\pi_{d,k}) \geq 0$ imply
\begin{equation}\label{SEP!}
    \frac{ \lambda_{\pi(1)} + \ldots + \lambda_{\pi(k)} }{k}  \geq
 \lambda_d - (d-k-1)\lambda_0\ ,
\end{equation}
for $k=1,\ldots,d-1\,$. In particular for $k=d-1$ one has
\begin{equation}\label{}
    \frac{ \lambda_{1} + \ldots + \lambda_{d-1} }{d-1}  \geq
 \lambda_d \ .
\end{equation}
We stress that conditions (\ref{SEP!}) are necessary but not
sufficient. For example if $d=3$ they give rise to
\begin{equation}\label{3I}
    \lambda_1,\lambda_2 \geq \lambda_3 - \lambda_0\ ,
\end{equation}
and
\begin{equation}\label{3II}
    \frac{\lambda_1 + \lambda_2}{2} \geq \lambda_3\ .
\end{equation}
In particular if $\lambda_0 \geq \lambda_3$ then (\ref{3I}) is
trivially satisfied and hence separability implies (\ref{3II}) which
is much weaker than (\ref{SEP}).

\begin{example} {\em Taking
\begin{equation}\label{}
    \lambda_0 = \ldots = \lambda_{d-1} = \frac{1-\lambda_d}{d} \ ,
\end{equation}
one recovers well known isotropic state
\begin{equation}\label{}
    \rho = \frac{1-\lambda_d}{d^2}\, \mathbb{I}_d \ot
    \mathbb{I}_d + \lambda_d P^+_d\ .
\end{equation}
Now, conditions (\ref{SEP}) and (\ref{SEP!}) coincide and give rise
to well known separability condition
\begin{equation}\label{}
    \lambda_d \leq \frac{1}{d+1}\ .
\end{equation}
Hence, our generalized class (\ref{FAM-G}) may be considered as a
simple generalization of the isotropic state. }
\end{example}

\section{Conlusions}

We provide full characterization of the family of 2-qudit states
defined in (\ref{FAM}). As a byproduct we introduce a 1-parameter
class of states which generalizes Horodecki construction in $d=3$
\cite{HOR}. It is shown that simple deformation of the original
family provided by (\ref{FAM-G}) is no longer `solvable', i.e. we
are not able to formulate complete list of necessary and sufficient
conditions for separability. It is hoped that our class of states
finds application in testing various notions in Quantum Information
Theory.

\section*{Acknowledgments}

This work was partially supported by the Polish Ministry of Science
and Higher Education Grant No 3004/B/H03/2007/33.

\end{document}